\documentclass[aps,prl,twocolumn,superscriptaddress]{revtex4}
\usepackage{graphicx}
\usepackage{color}

\newcommand{\ha}{$h_{\alpha}$}

\begin{document}

\title{Crediting multi-authored papers to single authors}

\author{Anna Tietze}
\affiliation{Institute of Neuroradiology, Charit\'e University Medicine Berlin, 13353 Berlin, Germany}
\email{annatietze@gmail.com}
\author{Serge Galam}
\affiliation{CEVIPOF - Centre for Political Research, Sciences Po and CNRS, 98 rue de l'Universite Paris, 75007, France}
\email{serge.galam@sciencespo.fr}
\author{Philip Hofmann}
\affiliation{Department of Physics and Astronomy, Interdisciplinary Nanoscience Center (iNANO), Aarhus University, 8000 Aarhus C, Denmark}
\email{philip@phys.au.dk}
\date{\today}
\begin{abstract}
A fair assignment of credit for multi-authored publications is a long-standing issue in scientometrics. In the calculation of the $h$-index, for instance, all co-authors receive equal credit for a given publication, independent of a given author's contribution to the work or of the total number of co-authors. Several attempts have been made to distribute the credit in a more appropriate manner. In a recent paper,  Hirsch has suggested a new way of credit assignment that is fundamentally different from the previous ones: All credit for a multi-author paper goes to a single author, the called ``$\alpha$-author'', defined as the person with the highest current $h$-index (\emph{not} the highest $h$-index at the time of the paper's publication) \cite{Hirsch:2019ab}. The collection of papers this author has received credit for as $\alpha$-author is then used to calculate a new index, \ha, following the same recipe as for the usual $h$ index. The objective of this new assignment is not a fairer distribution of credit, but rather the determination of an altogether different property, the degree of a person's scientific leadership. We show that given the complex time dependence of $h$ for individual scientists, the approach of using the current $h$ value instead of the historic one  is  problematic, and we argue that it would be feasible to determine the $\alpha$-author at the time of the paper's publication instead. On the other hand, there are other practical considerations that make the calculation of the proposed \ha~very difficult. As an alternative, we explore other ways of crediting papers to a single author in order to test early career achievement or scientific leadership. 
\end{abstract}
\maketitle

\section{Introduction}
\label{intro}

%multi-author credit distribution
Assigning appropriate credit to contributors of multi-authored publication is a challenging problem. Several schemes have been proposed to achieve this in a ``fair'' way  \cite{Tscharntke:2007aa,Schreiber:2008aa,Egghe:2008aa,Hirsch:2010aa,Galam:2011aa,Shen:2014aa,Vavrycuk:2018aa}. The issue has received special attention in connection with the $h$-index proposed by Hirsch in 2005 \cite{Hirsch:2005aa}, motivated by the huge importance that this single number has gained in the evaluation of scientists and institutions.  

%but it can be turned into a virtue - if the objective is to measure something else. The alpha-h
Recently, Hirsch has proposed an interesting variation to the $h$-index that uses a strongly biased credit distribution compared to all previous suggestions: Instead of giving (possibly normalized) credit to all co-authors, only a single author, the so-called $\alpha$-author, can claim credit for a given paper \cite{Hirsch:2019ab}. More precisely, the new index \ha~ is constructed in the same way as $h$ from a scientist's list of publications, but only publications are counted in which the person is the $\alpha$-author. The $\alpha$-author, in turn, is defined as the contributor with the highest value of the conventional $h$ at present. While \ha~ may at first seem to be even more ``unfair'' than $h$, it is not intended to serve the same purpose. Instead, it is meant to measure ``scientific leadership'', based on the assumption that the $\alpha$-author has contributed significantly to the conception and realization of the project leading up to the paper. In this sense, \ha~can be used to complement other, more conventional, measures of scientific success, including $h$ as such.

Hirsch's introduction of \ha~has quickly given rise to some criticism in Ref. \cite{Leydesdorff:2019aa}, partly because it was perceived to reinforce the Matthew effect in science, and partly because of a technical issue: In order to determine who the $\alpha$-author of a given paper is, Hirsch proposed to compare the $h$ values of all co-authors at the present time and not at the time of the paper's publication. This simplifies the calculation, as current $h$ values are readily available. In fact, it is often assumed that calculating historical $h$ values is difficult or impossible \cite{Leydesdorff:2019aa}. However, Leydesdorff \emph{et al.}  argued that using current $h$ values can lead to significant instabilities in the resulting of \ha's for collaborating scientists, since small relative changes in the $h$ values of the collaborators can lead to a shift of the $\alpha$-author status in many co-authored papers \cite{Leydesdorff:2019aa}. Hirsch has addressed both aspects of this criticism in Ref. \cite{Hirsch:2019aa}. 

The consequences of using the current $h$ values for the historic ones at the time of a paper's publication when calculating \ha~are, in fact, poorly understood and closely related to the time dependence of $h$ for individual scientists. In the present work, we show that historical $h$ values can be readily calculated using retrievable data from Web of Science and we study the time dependence of $h$ for a large number of condensed matter physicists. We show that  the average of $h$ over a larger population does indeed show a roughly linear time dependence, but this does not hold on the level of individual scientists. Given the relative ease of calculating historical $h$ values, the definition of \ha~could thus be modified such that the $\alpha$-author is determined at the time of a paper's publication. However, a more severe practical problem with \ha~is that it is extremely difficult to calculate due to (co-)author name disambiguation. 

As an alternative implementation of Hirsch's proposal, we consider variations of Galam's $gh$-index \cite{Galam:2011aa} to identify early career achievements or scientific leadership of authors from the position of their name in the author list. Such an approach is obviously only meaningful in a research field where the order of the authors encodes information about their contributions (as is the case in condensed matter physics). 

This paper is structured as follows: The Methods section briefly explains the data set used in this study and how to automatically calculate the time dependence of $h$ for a large group of scientists. In the Results and Discussion section we first present our results on the time dependence of $h$ and then discuss the possibility to use Galam's $gh$ to assign credit to single authors in order to investigate certain properties such as research leadership. We end the paper with a Conclusions section.

\section{Methods} 
\label{sec:1}

The data set used in this study has already been introduced in Ref. \cite{Tietze:2019aa} and it is described in detail there. In short, it consists of general citation data for 302 condensed matter physicists (number of publications, citations and $h$-values today) extracted from ResearcherID between April and December 2018, combined with detailed citation data for every paper co-authored by these individuals (24,286 papers), extracted from Web of Science (WoS). 

For a given individual from the group, the value of $h$ at any desired point in time can be determined as follows: In WoS, a search for all the individual's papers is performed using the Researcher ID number as unique identifier. Using the ``create citation report'' function, the details of every single paper can be obtained, i.e. the author list and the number of times the paper has been cited in each year after its publication. This data can be exported from WoS for further analysis. Calculating $h$ at a given point in time is now a trivial matter because it simply requires counting the number of published papers and the number of citations these papers have acquired up to that point in time. It is also possible to obtain the position of the author in question by inspecting the author list. This cannot be done entirely automatically, notably in the case of name changes or different spelling possibilities of a name. Note that the procedure offers a certain amount of protection against false Researcher ID profiles because publications which are not co-authored by the owner of the profile stand out.

The starting point $t=0$ is defined as the year a given author has published the first paper.  When a time dependence is considered, the starting year is also the year zero. When we refer to a 20 year career, we are thus considering years zero to 19.

\section{Results and Discussion} 
\label{sec:2}
\subsection{Time dependence of the $h$-index}

The time dependence of the $h$-index for individual researchers has been the central issue in the debate about the stability of the proposed \ha -index \cite{Hirsch:2019ab,Leydesdorff:2019aa,Hirsch:2019aa}. Determining the $\alpha$ author of a historical paper by choosing the author with the highest $h$ today is only unproblematic if $h$ increases linearly and with the same rate for all co-authors \cite{Hirsch:2019aa}. This assumption is obviously  questionable because a variation in the growth rates of $h$ is the feature that turns $h$ into a useful quantity in the first place. The time dependence of the $h$-index has been investigated previously  \cite{Egghe:2007ab,Egghe:2007ac,Burrell:2007aa,Guns:2009aa,Wu:2011aa,Mannella:2013aa,Tarasevich:2016aa}, albeit mostly theoretically or for a small number of individuals. Different curve shapes consistent with monotonic growth have been discussed \cite{Wu:2011aa}. To the best of our knowledge, no systematic study of the time dependence of $h$ for a well-defined larger group of scientists has been carried out so far.

\begin{figure}
 \includegraphics[width=0.5\textwidth]{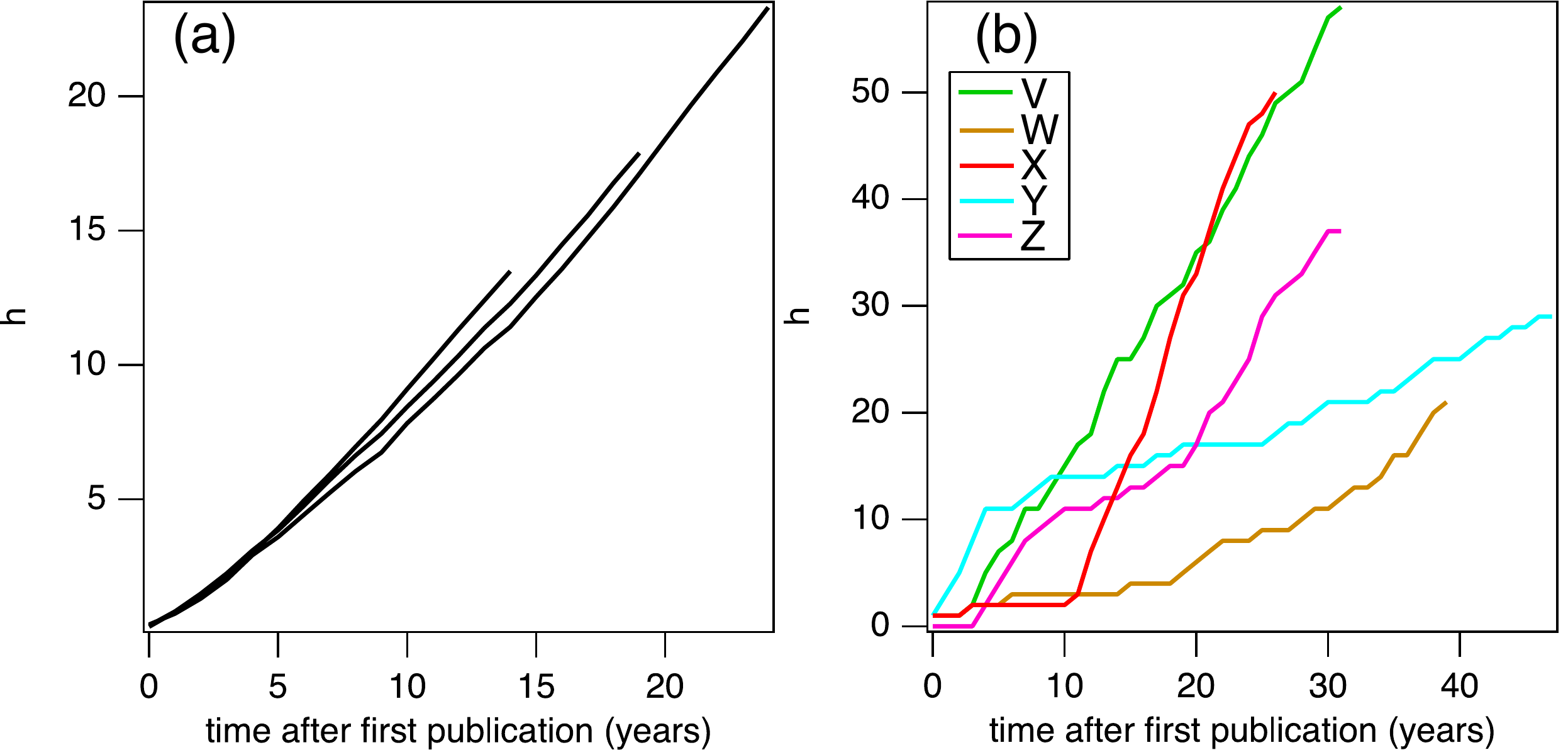}
\caption{(a) Time dependence for average $h$ for researchers with a career length of 15, 20 and 25 years. (b) Time dependence for five selected individuals with a qualitatively different behaviour. }
\label{fig:1}       % Give a unique label
\end{figure}

We start by investigating the average time dependence of $h$ for three sub-sets of the 302 individuals in our data set: scientists with a career length of up to 15 years (154 individuals), 20 years (105 individuals) and 25 years (67 individuals). Note that the group of 154 individuals with a career length of at least 15 years also contains all members of the other two groups.
The results are shown in Figure \ref{fig:1}(a). The assumption of a linear increase of $h$ with time is clearly reasonable, if not entirely correct. Remarkably, the slope of the curve is higher for the groups with a shorter career span. This can be expected because these groups also contain the  ``young'' researchers with a career start after 2002, and these profit especially from the general growth in the number of published papers and hence also citations \cite{Wuchty:2007aa}. 
% https://papers.ssrn.com/sol3/papers.cfm?abstract_id=3193712

%looking at the average is really misleading
A roughly linear time dependence of $h$ on average, however, does not imply that this also holds on the level of an individual researcher. In fact, this is not the case. Fig. \ref{fig:1}(b) illustrates some strong deviations from the linear behaviour for a few chosen examples from our data set. Researcher V shows almost ideal linear behaviour, apart from a slight delay in the start of the career. The other individuals show all types of different characteristics such an increased slope at later times (W and, rather extreme, X), or more complicated curve shapes (Y and Z). These different shapes and possible reasons for them have  been discussed previously \cite{Wu:2011aa}. It could be interesting to investigate this further by assigning researchers to different shape categories but this goes beyond the scope of the present work. Here, it is only important that a large variety of curve shapes is found and that their tendency to average out to a linear curve does not imply that linear behaviour holds on the level of individual researchers. With respect to the calculation of $h_{\alpha}$, a consequence of this is that from knowing $h$ today, it is impossible to make reliable statements about what $h$ has been at some point in the past. 
 % emphasize that this will be true for all that comes: whenever we look at an average, we have to keep this in mind. 

\begin{figure}
 \includegraphics[width=0.5\textwidth]{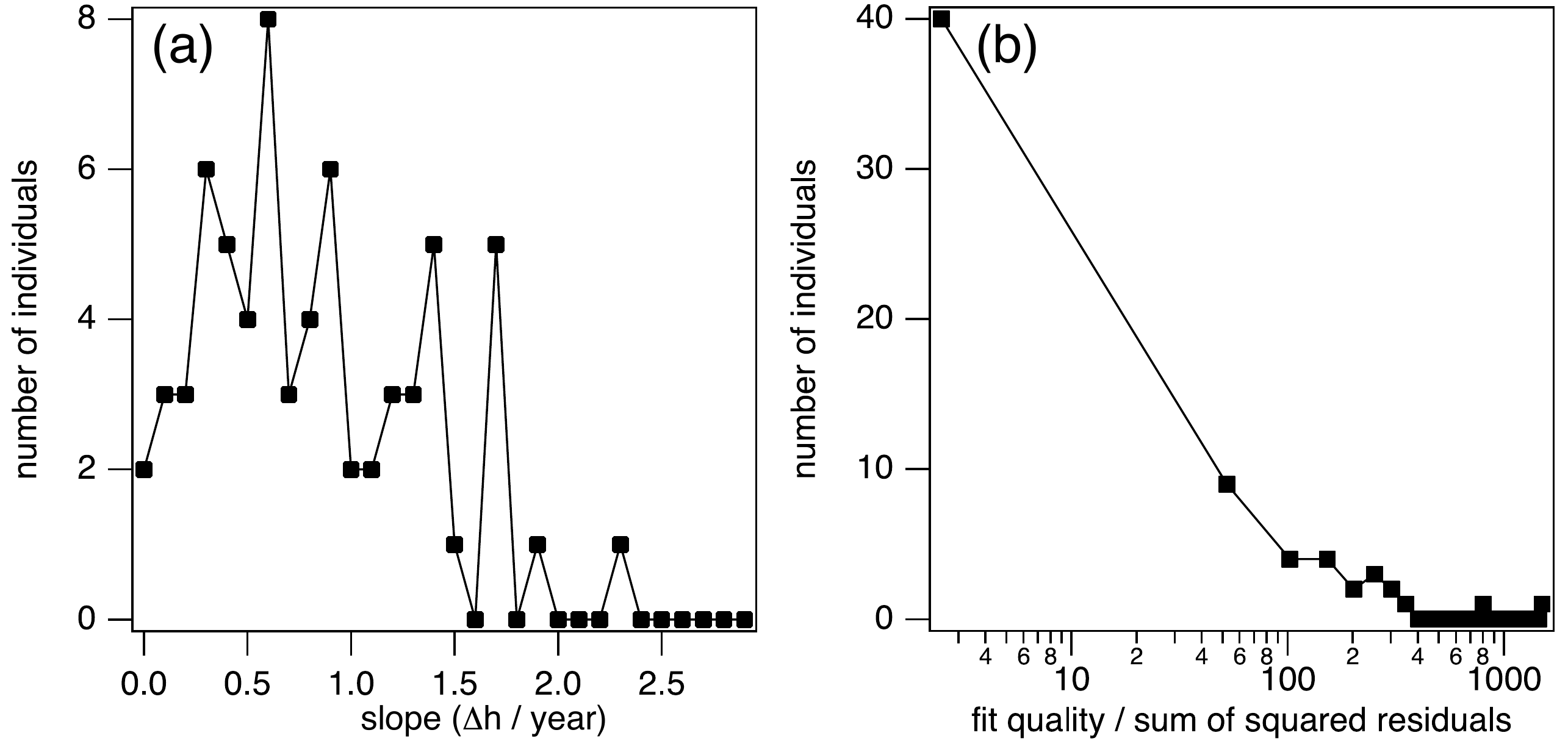}
\caption{Result of a linear fit to $h(t)$ for 67 individuals with a career length of 25 years or more, shown as histograms. (a) Slope and (b) goodness of linear fit (expressed as the sum of squared residuals). }
\label{fig:2}       % Give a unique label
\end{figure}

% quantify deviation from average
In order to quantify the deviation from a linear time dependence, we perform a linear fit  of  $h(t)$ for individuals of the sub-group with a career length of at least 25 years, using the constraint that $h$ in the first year of the career $h(0)$ matches the actually observed value (mostly 0, but sometimes 1 or even 2).  The resulting slope and the sum of the squared residuals are shown as histograms in Figure \ref{fig:2}(a) and (b), respectively. We see that $h$ typically increases at a rate of 0-2 per year and that, while a linear fit works reasonably well in most cases, there are many individuals for which it does not. 

% what happens when somebody dies?
Using the current instead of the historical value of $h$ to determine the $\alpha$-author of a paper has another potential drawback: If one of the authors of a paper ceases to publish, he / she could be ``out-$\alpha$'ed'' by the others who continue doing so (except for especially outstanding authors with a very high $h$).
To illustrate this, let us assume that three collaborators (A, B, C) with similar $h$ values have published many papers together. At some point, B and C stop publishing due to a change of career or owing to certain life circumstances. This leads to B's and C's $h(t)$ to level off while growth continues for A. Eventually, A would out-$\alpha$ B and C for reasons that are not related to scientific leadership.  How long would A need to wait until his / her $h$ value would overtake those of B and C? 

\begin{figure}
\includegraphics[width=0.5\textwidth]{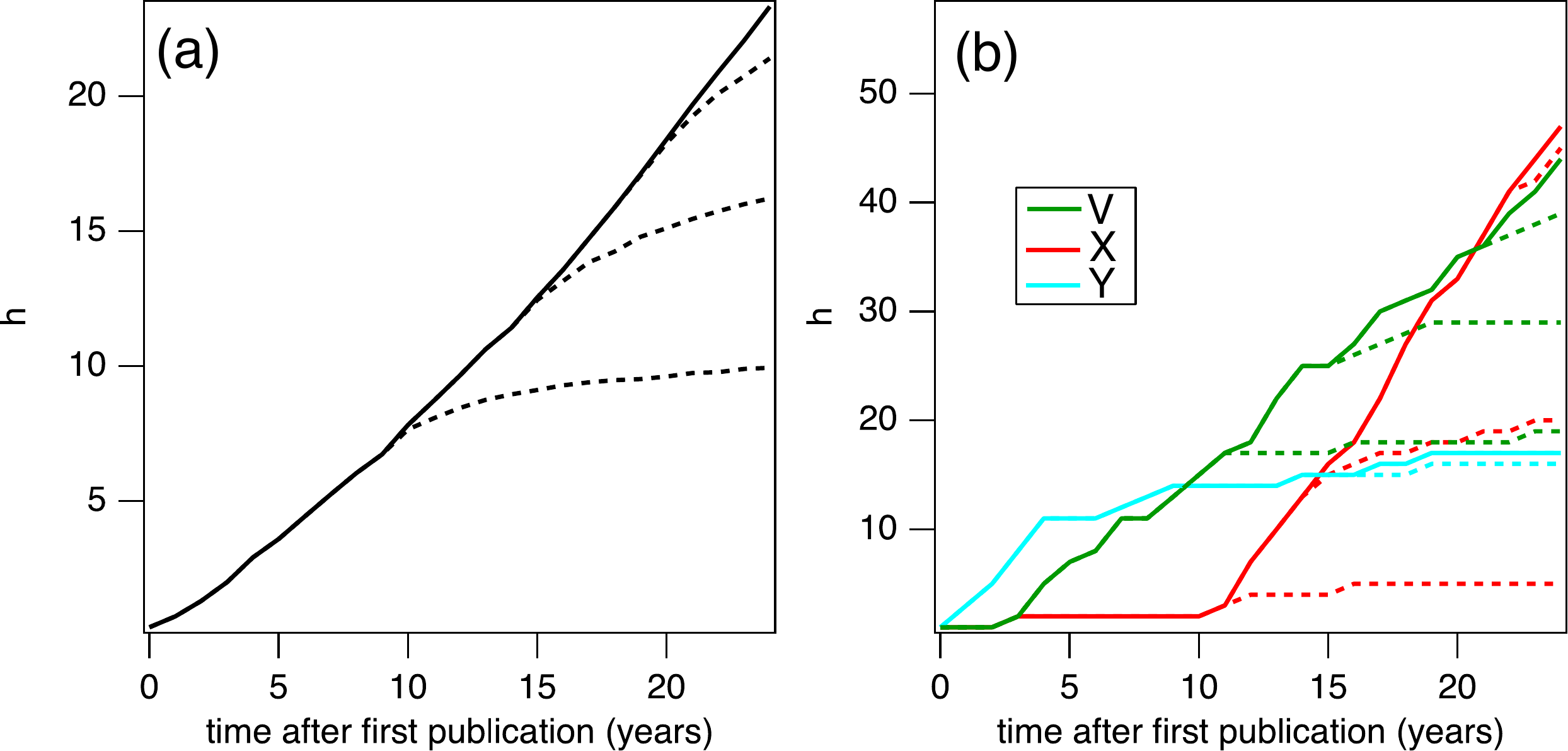}
\caption{Time dependence of $h$ when an end to publication activity is artificially enforced. (a) Average $h(t)$ for 67 individuals with a research career length of at least 25 years (solid line). The dashed lines show the $h(t)$ when papers published after the first 10, 15 and 20 years are not considered. (b) Corresponding curves for three of the 5 researchers from Figure \ref{fig:1}(b) (only three are chosen for clarity.}
\label{fig:3}       % Give a unique label
\end{figure}\

The saturation of $h$ following the end of a scientific career has already been discussed by Hirsch in his first paper on the subject \cite{Hirsch:2005aa}. Using a simple model, he pointed out that the time needed for $h(t)$ to level off increases with the total length of the scientific career / the total number of papers published. Saturation in Hirsch's model results from all papers ending up in the $h$-core, and while this is not expected in a realistic scenario, the key-feature of a slower saturation after a long career is still found in our data.  Fig. \ref{fig:3}(a) shows the average $h(t)$ for the group of scientists with a career length of 25 years or more as a solid line, as already given in Fig. \ref{fig:1}(a). The dashed lines show the resulting curve if we artificially enforce a career end after 10, 15 or 20 years. The expected tendency for $h(t)$ to level off is indeed observed and this appears to happen faster after a shorter career. 

It is important to remember that, again, the trend observed for the average population does not permit conclusions about the trend for individual researchers, which can be very different. This is illustrated in Fig. \ref{fig:3}(b) which shows the corresponding curves for researchers V, X and Y from the data in Fig. \ref{fig:1}(b). In this case, only the curve for V is similar to the (roughly linear) average whereas the curves for X and Y and quite different.   

An important conclusion from this section is that determining the $\alpha$-author on the basis of the current $h$ values does not seem to be a good choice. An obvious solution to the issue would be to determine the $\alpha$-author of a paper by using the historical $h$ values at the time of the paper's publication. Leydesdorff \emph{et al.} argue that this is challenging because the required citation data are not provided directly in WoS  \cite{Leydesdorff:2019aa} but, as we have shown here, they can actually be extracted -- provided that the author's paper collection can be identified without ambiguity by a search in WoS. 

The biggest practical difficulty in the calculation of $h_{\alpha}$ is, in fact, another one: Even if all information about every article authored by a person, including detailed citation data, is available, this is not sufficient. The same information is needed for every co-author on every paper the person has ever published with and this is essentially impossible to obtain, unless all authors have unique identifiers. Such unique identifiers are being currently introduced (\emph{e. g.} ORCID or Researcher ID) but even if such identifiers were used throughout today (which they are not), it would still take at least 20 years before a calculation of $h_{\alpha}$ would be practical. 

\subsection{Alternative approaches to crediting publications to single authors}

As we have shown in the previous section, the proposed $h_{\alpha}$ suffers from two important drawbacks: (1) Using the current values of $h$ as a proxy for the historical values of $h$ in order to determine the $\alpha$-author of a paper causes several problems. (2) Mainly because of name disambiguation, determining $h_{\alpha}$ is extremely difficult and can probably only be done for authors one is very familiar with, or who have sufficiently high $h$ values, such that all past and current competitors for the $\alpha$-status can be identified and checked by person familiar with the field of research (this does not protect against situations like former PhD students who have meanwhile obtained a high $h$ value through research in a different field). The first problem for the calculation of \ha~can be fixed quite easily but the second can not.
Still,  the idea of an ``unfair'' credit distribution to a single author in order to identify characteristics such as research leadership is very interesting and in this sub-section, we explore alternative, more practical, approaches to accomplish this. 

A useful tool for giving credit to a single author only is the $gh$-index proposed by Galam \cite{Galam:2011aa}. Originally, this was introduced as an index that obeys conservation laws for the number of published papers and citations. Indeed, the current practice of all co-authors taking credit for the entire paper and all of its citations appears to violate elementary conservation laws. As an illustration, imagine a business where every associated partner gets the total business revenue and also owns 100\% of the business. Unfortunately, this is too good to be true and only feasible with one single owner. A soon as two or more partners own the business, the profit and the ownership must be divided along the respective shares of partners whose total equals 100\% of  the business assets. Similarly any meaningful quantitative bibliometrics treatment should obey the same kind of conservation law, here with respect to numbers of papers and citations. 

While the $gh$-index was proposed with the intention of achieving a ``fair'' distribution of credit, it can be defined in a rather large number of declinations, each one giving a specific distribution of a paper's citations to each co-author. This distribution can also be chosen to be extremely ``unfair'', such as attributing all of a paper's citations to a single author with the purpose of measuring quantities different from the usual total impact. We thus define the following variations of $gh$:

\begin{itemize}
\item $gh_{e}$ gives equal credit to every author  by dividing the number of a paper's citations by $k$ if there are $k$ authors, and then follows the same recipe as for the calculation of $h$. $gh_{e}$ can thus be seen as a variation of the usual $h$-index but adapted to multi-author publications and conserving the total number of citations.
\item $gh_{1}$ gives all credit to the first author, i.e. all citations of a paper are counted if the author is the first in the author list and otherwise the citations are set to zero. This is again followed by the usual procedure of calculating $h$. $gh_{1}$ can be viewed as an index for early career achievements, or  in connection with  single author publications.
\item $gh_{L}$ gives all credit to the last author and is otherwise calculated in the same way as $gh_1$.  In condensed matter physics, as in many other fields, the last author is usually the person with the overall responsibility for a project. In the sense of Hirsch's suggestion, this would be the $\alpha$-author but it would not necessarily be the person with the highest values of $h$. 
\end{itemize}

\begin{figure*}
\includegraphics[width=1.0\textwidth]{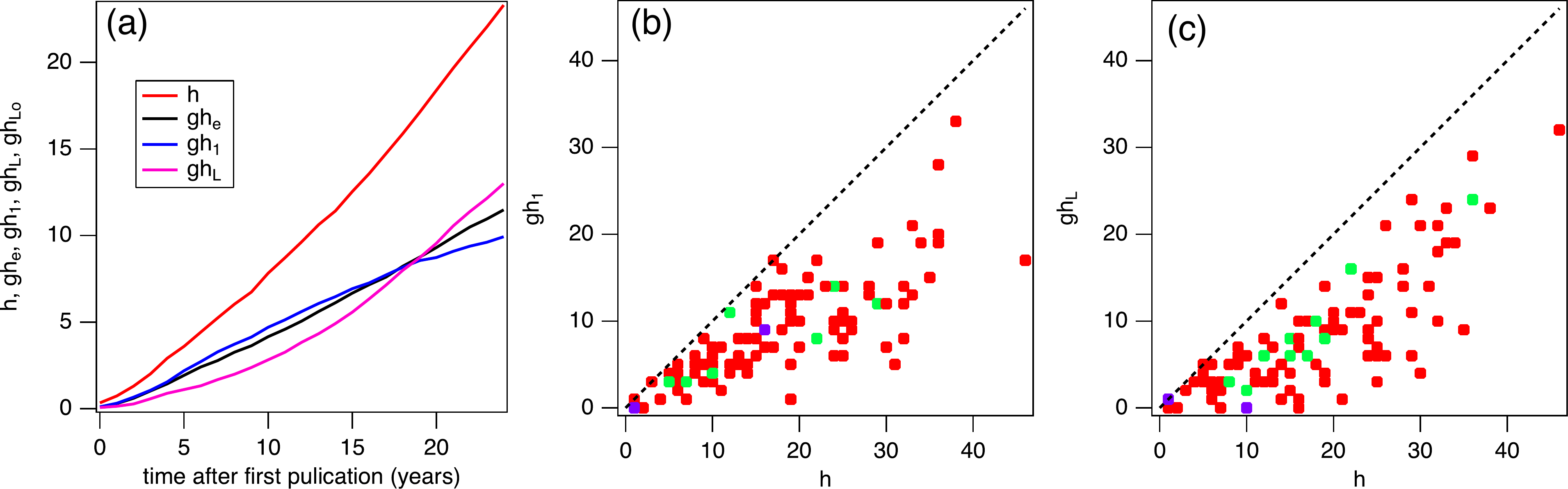}
\caption{(a) Average time dependence of $h$, $gh_e$, $gh_1$ and $gh_L$ for the sub-group of researchers with a career length of at least 25 years. (b) and (c) Distribution of $gh_1$ and $gh_L$ as a function of $h$ for the 105 researchers with a career length of at least 20 years. Multiple incidents are colour-coded as red (1), green (2) and purple (3).  }
\label{fig:4}       % Give a unique label
\end{figure*}

If probing research leadership is the objective, using $gh_{L}$ is much simpler than using \ha~because its calculation is based on the easily determined position of an author in the list of co-authors. An important restriction is that the author position needs to contain information about author's role in a collaboration which is not always the case. If applicable in a given field of research, choosing the last author instead of the one with the highest $h$ as the leading author can have several advantages. Some have already been mentioned, for example the presence of high-$h$ co-authors on a paper who are simply on the author list because they have contributed to the funding of the project \cite{Leydesdorff:2019aa}, a practice that is not desirable but common. Another example could be a collaboration involving several groups, all lead by senior people with a high $h$. Even if the project is lead by one individual, this would not necessarily be the person with the highest $h$, especially if the groups cover different sub-disciplines. In condensed matter physics, for instance, density functional theory is a sub-field of huge impact and therefore high citation numbers and $h$-indices \cite{VanNoorden:2014aa}. A senior individual working in this field would be likely to out-$\alpha$ the other co-authors.

Note that both $gh_1$ and $gh_L$ contain single author papers. If $gh_L$ is to be used as an indicator of research leadership, it is not obvious that single author papers should be included. On the other hand, single author papers are quite rare (less than 3\% of the total for our data set), and we can therefore ignore this issue here.

The average of the three $gh$ indices is plotted in Fig. \ref{fig:4}(a) for the authors with career length of at least 25 years and compared to the average $h$ for the same group. All $gh$'s are smaller than $h$ because they do either include a normalization ($gh_{e}$) or represent only a subset of an author's publications ($gh_1$ and $gh_L$). As one might expect, $gh_{e}$ is somewhat similar to $h$, in that it is almost linear but has a slightly increasing slope over time.  $gh_{1}$, on the other hand, shows a decreasing slope over time. This might be expected because many first author papers are written in the beginning of a researcher's career and the total number of citations they receive saturates at a later career stage. This is completely equivalent to the case of ceasing to publish altogether, just restricted to first author papers. 
The overall characteristics of $gh_{L}$ with a clearly increasing slope over time is also consistent with what would be expected for a $h$-type index associated with research leadership. Last author papers are typically first published with the beginning of an independent career and their number and impact  first becomes apparent at a later career stage.

As in the case of the $h$-index, merely inspecting the average of $gh_1$ and $gh_L$ is insufficient to draw conclusions on the level of a single researcher. In Figure \ref{fig:4}(b) and (c), we therefore show the distribution of $gh_1$ and $gh_L$ as a function of the conventional $h$, respectively. In the case of $gh_1$, we observe that the role of first author papers decreases for individuals with a high value of $h$. This is to be expected because a high $h$ is mostly based on collaborative work with changing first authors. The distribution of $gh_L$ is more interesting, especially when it comes to using this as a possible indicator of research leadership. The distribution of $gh_L$ values for a given $h$ is, in fact, very broad. For $h=29$, for example, $gh_L$ varies between 6 and 24 meaning that one author has achieved a relatively high value of $h$ with only 6 last author publications while another has done the same with 24. This large variation clearly indicates that $gh_L$ could be a useful quantity to complement $h$. For the two authors in question, one would conclude that both made a significant contribution to their research field with the latter in a leading role but not the former. 

\section{Conclusion}

We have inspected the time dependence of the $h$-index for a large number of individuals. While the average is roughly a linear function of time, this does not hold on the level of individual researchers and we have discussed the implications of this finding for the calculation of the \ha-index recently suggested by Hirsch. Based on our findings, it appears more appropriate to identify the $\alpha$-author of a paper based on the historical $h$-values at the time of the paper's publication rather than on the current ones, and we have shown that this is relatively easy. On the other hand, there are other severe practical limitation for a calculation of \ha. We have adapted Hirsch's suggestion to assign credit in a multi-author paper to a single author in $h$-type indices for other purposes than merely identifying total research impact,  and we have implemented this using the scheme of the $gh$-index suggested by Galam. Such $h$-type indices could, for instance, be used to study the correlation between early career success (\emph{e. g.} a high $gh_1$ or many first author papers in the initial career stage) and later career achievements (such as a high $h$ or $gh_L$).

\section{Acknowledgements}
This work was supported by VILLUM FONDEN via the center of Excellence for Dirac Materials (Grant No. 11744).  One of us (SG) would like to thank J.~E.~Hirsch for helpful discussions.

%
%\bibliography{Hindex}   % name your BibTeX data base
%

\end{document}